\documentclass[12pt,leqno]{article}
\usepackage{amsmath,amsthm}

\def\init{\setcounter{equation}{0}}
\newtheorem{theorem}{Theorem}[section]
\newcommand{\R}{{\bf R}}

\newcommand{\Z}{{\bf Z}}

\newtheorem{pro}{Proposition}[section]

\newcommand{\e}{{\varepsilon}}

\usepackage{tikz}

\title{A simple proof of magnetic and electric Aharonov-Bohm effects
\author{G.Eskin, \ \ \  Department of Mathematics, UCLA,\\ Los Angeles,
CA 90095-1555, USA. \ E-mail: eskin@math.ucla.edu}
}

\begin{document}

\maketitle

\begin{abstract}
Magnetic Aharonov-Bohm effect (AB effect)  was studied in hundreds of
papers starting with the seminal paper of Aharonov and Bohm [AB] published in 1959. 
We give a new   proof of the magnetic  Aharonov-Bohm effect
without using the scattering theory and the theory of inverse boundary value
problems. 
We consider separately the cases of one and several obstacles.  
The electric  AB effect was studied much less.
We give  the first  proof of the electric AB effect in domains with moving boundaries.
When the boundary does not move with the time the electric AB effect is absent.  
\end{abstract}

\section{Introduction.}
\label{section 1}
\init

Let $\Omega_1$  be a bounded domain in $\R^2$,  called the obstacle.  
  Consider the time-dependent 
 Schr\"odinger equation  in $(\R^2\setminus\overline \Omega_1)\times (0,T)$:
\begin{equation}                                       \label{eq:1.1}
-ih\frac{\partial u}{\partial t}+
\frac{1}{2m}\sum_{j=1}^n\left(-ih\frac{\partial}{\partial x_j}
-\frac{e}{c}A_j(x)\right)^2u+e V(x)u=0,
\end{equation}
where  $n=2$,
\begin{equation}                                           \label{eq:1.2}
u|_{\partial \Omega_1\times (0,T)}=0,
\end{equation}
\begin{equation}                                           \label{eq:1.3}
u(x,0)=u_0(x),\ \ x\in \R^2\setminus\Omega_1,
\end{equation}
and the electromagnetic potentials $A(x), V(x)$ are independent of $t$.
The equation (\ref{eq:1.1}) describes an electron in $(\R^2\setminus\overline \Omega)\times(0,T)$,
more precisely,  $|u(x,t)|^2$  is a probability density of finding the electron in an infinitesimal
 neighborhood of $x$.
Let
\begin{equation}                                           \label{eq:1.4}
\alpha=\frac{e}{hc}\int_\gamma A(x)\cdot dx
\end{equation}
be the magnetic flux,  where $\gamma$  is a simple closed contour containing $\Omega_1$.
In  the   seminal paper [AB]  Aharonov  and  Bohm discovered  
that even when the magnetic field $B(x)=\mbox{curl\,}A=0$
in $\R^2\setminus \overline \Omega_1$,    
the magnetic potential $A$  affects the electron in 
$\R^2\setminus\overline \Omega_1$  if  $\alpha\neq 2\pi n$,   $\forall n \in Z$.
  This phenomenon is called the Aharonov-Bohm  effect,  where we say that the Aharonov-Bohm effect takes 
place if there are solutions $u(x,t)$  of (\ref{eq:1.1})  for which a physical
quantity such as the probability  
density $|u(x,t)|^2$  or probability current 
$\mbox{Re\,}(\overline{u(x,t)}(-i\frac{\partial}{\partial x})u(x,t))$
depends on the gauge equivalence class of the magnetic potential.

The purpose of the present paper is to give a simple proof of the AB effect without using
the scattering theory or the theory of inverse boundary value problems.

Aharonov and Bohm proposed 
a physical experiment  to test the AB effect.  The experimental proof of AB effect  was
not easy to achieve
in the way that is free of any controversy.  This  was 
accomplished by Tonomura 
et al [T et al].  
In the same paper [AB] Aharonov and Bohm gave a mathematical proof of AB effect 
in the case 
when 
the obstacle is reduced to a point:  $\Omega_1=\{0\}$.
They explicitly constructed the scattering amplitude and showed that the scattering 
cross-section is influenced by the magnetic flux modulo $2\pi n,\ n\in \Z$.
Later Ruijsenaars [R]
proved the same result in the case when the obstacle is a circle.  The next important
step was done by Nicoleau [N] who found a way to compute the integrals
\begin{equation}                                         \label{eq:1.5}
\exp\Big(\frac{ie}{hc}\int_{-\infty}^\infty A(x_0+t\omega)\cdot\omega dt\Big) 
\end{equation}
and
\begin{equation}                                         \label{eq:1.6}
\int_{-\infty}^\infty V(x_0+t\omega)dt,
\end{equation}
knowing 
the scattering operator.  In (\ref{eq:1.5}), (\ref{eq:1.6}) $n=2$  or $n=3$  and \linebreak
$x=x_0+t\omega,\ t\in \R$,  is any straight line that does not intersect  the obstacles.  
When the obstacle $\Omega$ is convex,  Nicoleau [N]  used (\ref{eq:1.5}),
(\ref{eq:1.6}) and the $X$-ray transform  to determine $B=\mbox{curl\,}  A, \ V(x)$  and the
magnetic flux $\alpha(\mbox{mod\, } 2\pi m)$ for $n=2$, i.e. he solved the inverse scattering problem.     

When considering the Aharonov-Bohm effect,
the magnetic field  \linebreak $B(x)=0$, and therefore one needs only 
a finite number of integrals of the form (\ref{eq:1.5})  to  
determine  the magnetic fluxes modulo $2\pi p,\ p\in \Z$.
In this case the $X$-ray  transform  is not needed.  Still  the recovery 
of integrals (\ref{eq:1.5})  from the scattering
operator is quite complicated. 
The motivation of the present work was to find an easier proof of the AB effect 
that does not involve the scattering theory.

In [W1] 
Weder considered the case 
when
$n=2$  and $V(x)=0$  and he recovered integrals (\ref{eq:1.5})
from the scattering  operator in a larger  class of potentials then in [N]. 
  He also considered the case when the obstacle
is reduced to a point as in the original Aharonov-Bohm paper [AB].

In a series of papers [E3], [E4], [E5], [E6] the inverse boundary value problems
for the Schr\"odinger equations with time-independent
electromagnetic potentials were studied.
The most complete result was obtained in [E6] where an arbitrary number of obstacles 
(not necessary convex) was considered, and 
it was proven that the boundary data determine the gauge equivalent class
of electromagnetic potentials.  In particular,
the magnetic AB effect was proven.  In [E6]  and [E3]  the case of
Yang-Mills potentials generalizing electromagnetic potentials was also 
considered.

It is well known  that the inverse boundary value problem can be reduced to 
 the inverse scattering problem, and vice versa,  in the case when the magnetic
and electric potentials have a compact support.

In the case of the magnetic AB effect it is natural 
to assume that the magnetic field $B(x)$  has a compact support.
The electric potential does not affect  the magnetic AB effect  and,
for the simplicity,  we can assume that $V(x)$ has 
a compact support too,
or even $V(x)$ equal to zero.
If $B(x)$  has a compact support and  
 if $n\geq 3$,  or $n=2$  and the total
 magnetic flux is zero, then  one can find a magnetic potential 
$A(x)$  with compact support such that $B(x)=\mbox{curl\,}A(x)$  (cf. [E4], \S1).
In this case the results of [E6]  lead to the solution of the inverse scattering
 problem and, in particular,  to the proof  of the magnetic AB  effect.

In [BW1] Ballesteros and Weder 
recovered integrals  (\ref{eq:1.5}), (\ref{eq:1.6})  from the scattering
operator for $n=3$  in a  class of potentials larger than in [N]  with less 
requirements on the smoothness.  They also proved in [BW1]
 the AB effect for some 
quite restrictive class of 
obstacles in $\R^3$. 
When the electric potential has a compact support their results were preceded by [E6]
where a general class of obstacles was considered.

When $n=2$  and the total flux is not zero  the magnetic potential is $O(\frac{1}{|x|})$  
even when $B(x)$  has a compact support.  In this case the inverse scattering problems
become more difficult.  This case was treated in [EIO] (see also [EI]).
The proofs in [EIO], [EI] use results of [N] (or [W1])  and also
the results of Yafaev and Roux-Yafaev ([RY1], [RY2], [Y])  on the singularities 
of the scattering matrix in the case of long range magnetic potentials.

All previous papers  
considered  the case of electromagnetic potentials independent of the time.
There are only two papers [E1]  and [W3] that consider inverse problems
for the Schr\"odinger equations with time-dependent potentials but [W3] has
no relation to the AB effect and
[E1]
 is the only  paper studying the inverse boundary value problem for
the time-dependent electromagnetic potentials. 
 In particular, [E1] gives the proof of 
the combined electric and magnetic AB effect.

Now  we shall describe the content of the present paper.

We start in \S3 (subsection 3.1) with the proof of the magnetic AB effect  
in the case of one obstacle in $\R^2$  by constructing solutions of nonstationary
Schr\"odinger equations depending  on a large parameter.
We construct such solutions in two steps:  first we construct a geometric optics 
solutions  
of the wave equation concentrated in a small neighborhood  of a ray  and
then we use the Kannai's formula (cf. [K])  that transform the solution of
the wave equation into the solution of the Schr\"odinger  equation.
We use these solutions to prove the following theorem
\begin{theorem}                                  \label{theo:1.1}
There exists a solution $u(x,t)$  of the Schr\"odinger equation (\ref{eq:1.1})
in $(\R^2\setminus\overline\Omega_1)\times(0,T)$  with the boundary condition (\ref{eq:1.2}) and
some highly oscillating initial data (\ref{eq:1.3}) such that 
\begin{equation}                                     \label{eq:1.7}
|u(x,t)|^2 = 2\sin^2\frac{\alpha}{2}+O(\e)
\end{equation}
in an $\e$-neighborhood of some point,  where $\e$ is small and $\alpha$ is the magnetic flux
(\ref{eq:1.4}), $\alpha\neq 2\pi n,\ \forall n\in\Z$.  Therefore,  the probability density $|u(x,t)|^2$  depends
on the magnetic flux $\alpha$, i.e. the AB effect holds.  Moreover $|u(x,t)|^2$  determines
$\alpha(\mbox{mod\ } 2\pi n,\ n\in \Z)$  up to a sign.
\end{theorem} 

In a short subsection 3.2 we extend Theorem \ref{theo:1.1} to the three-dimensional case.

In subsection 3.3  we prove a theorem similar to Theorem \ref{theo:1.1}   in the case of several obstacles  
in $\R^2$  (see Theorem \ref{theo:3.1}  below). The case of  several obstacles  
requires the construction of the geometric optics solutions for the wave equation 
concentrated in the neighborhood of broken rays,  i.e. rays  reflected at the 
obstacles.

The use of broken rays allows to treat the case 
when obstacles are close to each other and the treatment of the cluster of
obstacles as one obstacle  may miss the AB effect. It is mentioned 
in the Remark 3.1  that the broken rays are used   sometimes in the case of one 
obstacle too.

In \S 4  the electric AB effect is considered,  assuming  that the magnetic 
potential
$A=0$  and  the electric field $E=\nabla V(x,t)=0$  in 
the domain $D\subset \R^n\times[0,T]$.  The electric 
AB effect is  studied much less than the magnetic AB effect.  When  the domain 
$D$  has  the form $\Omega\times (0,T)$,  where $\Omega\subset\R^2$  is connected,
then $E=\nabla V(x,t)=0$  in $D$ implies that $V(x,t)=V(t)$.
Such electric potentials are gauge equivalent to zero electric potential,  i.e.
there is no electric AB effect.  It is not surprising  that there 
was  neither mathematical  nor experimental proof of electric AB effect in
such domains.  To get an electric AB effect one needs to consider domains $D$ such 
that the topology of the intersection $D\cap\{t=t_0\}$ changes 
with $t_0$,  i.e. the boundary of $D$ moves with the time.  Such domains are considered
in \S 4 and the following theorem is proven
\begin{theorem}                                                 \label{theo:1.2}
Let $u(x,t)$  be the solution of the Schr\"odinger equation
\begin{equation}                                           \label{eq:1.8}
ih\frac{\partial u}{\partial t}+\frac{h^2}{2m}\Delta u - e V(x,t)u=0,\ \ \ (x,t)\in D
\end{equation}
with zero boundary conditions and nonzero initial condition.  Let $v(x,t)$  be the solution of (\ref{eq:1.8})
with $V(x,t)=0$  and the same initial  and boundary conditions as $u(x,t)$.  Suppose that
$E=\frac{\partial V(x,t)}{\partial x}=0$  in $D$ and that $V(x,t)$  is not gauge equivalent 
to zero potential.  Then there exists a domain  $D$  such that the probability densities
$|u(x,t)|^2$  and $|v(x,t)|^2$ differ in $D$,  i.e.  the impact of the electric potential
$V(x,t)$  is different from the impact of the zero potential.

This proves the electric AB effect.
\end{theorem}

\section{The magnetic AB effect}
\label{section 2}
\init

Let $\Omega_1,...,\Omega_m$  be smooth obstacles in $\R^n$.  Assume 
that
$\overline \Omega_j\cap\overline \Omega_k\neq 0$ if $j\neq k$.
Consider the  Schr\"odinger equation (\ref{eq:1.1})  in 
$(\R^n\setminus\overline\Omega)\times (0,T), \ n\geq 2$,  where 
$\Omega=\cup_{j=1}^m\Omega_j$,
\begin{equation}                                 \label{eq:2.1}
u\big|_{\partial\Omega\times(0,T)}=0,
\end{equation} 
and (\ref{eq:1.3}) holds in $\R^n\setminus\Omega$.

Let $B(x)=\mbox{curl\,}A(x)$  be the magnetic field,  $n=2$ or 3.
In the case $n>3$  we consider $B(x)$  as the differential of the form 
$\sum_{j=1}^n A_j(x)dx_j$.

In this paper we assume that $B(x)=0$  in $\R^n\setminus\overline\Omega$,  
i.e.  the  magnetic field $B(x)$ is shielded inside $\Omega$.  For  the simplicity, 
we assume  that $V(x)$  has a compact support.

Denote by  $G(\R^n\setminus \Omega)$  the group of $C^\infty$ complex-valued
functions $g(x)$ such that $|g(x)|=1$ in $\R^n\setminus\Omega$ 
and
\begin{eqnarray}
\nonumber
g(x)=1+O\Big(\frac{1}{|x|}\Big)\ \ \mbox{for}\ \ |x|>R\ \ \mbox{if}\ \ n\geq 3,
\\
\nonumber
g(x)=e^{ip\theta(x)}\bigg(1+O\Big(\frac{1}{|x|}\Big)\bigg)\ \ \mbox{for}\ \ |x|>R\ \ 
\mbox{if}\ \ n=2.
\end{eqnarray}
Here $p$ is an arbitrary integer, $0\in\Omega$  and $\theta(x)$  is
the polar angle  of $x$.  
The difference between the cases $n=2$  and $n\geq 3$  is 
the consequence of the fact
that the set $\{x:|x|>R\}$  is 
simply connected when $n\geq 3$ and it is not simply-connected when  $n=2$.
We call
$G(\R^n\setminus \Omega)$  the gauge group.   If  $u'(x,t)=g^{-1}(x)u(x,t)$  then
$u'(x,t)$  satisfies the Schr\"odinger
equation (\ref{eq:1.1})   with electromagnetic potentials $(A'(x),V'(x))$,  where
$V'(x)=V(x)$  and
\begin{eqnarray}                                       \label{eq:2.2}
\frac{e}{c}A'(x)=\frac{e}{c}A(x)+ihg^{-1}(x)\frac{\partial g(x)}{\partial x}.
\end{eqnarray}
We shall call magnetic potentials $A'(x)$  and  $A(x)$  
gauge equivalent   if there exists $g(x)\in G(\R^n\setminus\Omega)$
such that (\ref{eq:2.2})  holds.

We shall describe all gauge equivalence classes  
of 
magnetic potentials when  $B=\mbox{curl\,}A=0$  in  $\R^n\setminus\Omega$.

It is easy to show (see, for example, \S 4  in [E1])  that $A(x)$ and $A'(x)$  
are gauge equivalent iff for any closed contour $\gamma$  in $\R^n\setminus\overline\Omega$  we have
$$
\frac{e}{hc}\int_\gamma A(x)\cdot dx - \frac{e}{hc}\int_\gamma A'(x)\cdot dx = 2\pi p,
$$
where $p\in\Z.$

Since $\mbox{curl\,}A=0$  in $\R^n\setminus\overline\Omega$  the integral $\int_\gamma A(x)\cdot dx$
depends only on homotopy class of $\gamma$  in $\R^n\setminus\Omega$.

Consider     the case of one obstacle $\Omega_1$  in $\R^2$.  
Let $\gamma_1$  be a simple closed contour in $\R^2\setminus\overline\Omega_1$ containing 
$\Omega_1$.  Any closed contour $\gamma$ in $\R^2\setminus\overline\Omega_1$
is homotopic to $m\gamma_1$,  where $m\in\Z$. Therefore the gauge equivalent class in 
$G(\R^2\setminus\Omega_1)$ is determined by the magnetic flux
$\frac{e}{hc}\int_{\gamma_1} A(x)\cdot dx$  modulo $2\pi p,\ p\in\Z$.

In the case of several obstacles $\Omega_1,...,\Omega_m$  in $\R^2$  denote
by $\gamma_j,\ 1\leq j\leq m,$  a simple closed curve encircling $\Omega_j$  only.
Let
$$
\alpha_j=\frac{e}{hc}\int_{\gamma_j}A\cdot dx
$$
be the corresponding  magnetic flux. 

Any closed  contour $\gamma$  in $\R^2\setminus\bigcup_{j=1}^m\overline\Omega_j$  is
homotopic to $\sum_{j=1}^m n_j\gamma_j,\linebreak n_j\in \Z$.  Therefore the numbers 
$\alpha_j(\mbox{mod\,}2\pi p_j),\ p_j\in Z,j=1,...,m$,  determine the gauge equivalence class of $A(x)$.
Analogously,  in the case  $m\geq 1$  obstacles in $\R^n,\ n\geq 3$,  any closed contour in
$\R^n\setminus \overline\Omega$   is homotopic to $\sum_{j=1}^r n_j\beta_j$,  
where $\{\beta_1,...,\beta_r\}$
is the basis of the homology group of $\R^n\setminus\overline\Omega,\ n_j\in\Z$.  Therefore
$A(x)$  and $A'(x)$  are gauge equivalent iff
$$
\frac{e}{hc}\int_{\beta_j} A(x)\cdot dx - \frac{e}{hc}\int_{\beta_j} A(x)\cdot dx =2\pi p_j,\ p_j\in\Z,
$$
for all $1\leq j\leq r$.
\qed

Any  two  magnetic potentials belonging  to the same gauge equivalence class 
represent the same physical reality  and can not be distinguished in any physical experiment.

 Consider   the probability density $|u(x,t)|^2$.
It has the same value for any representative of the same gauge equivalence class since 
$|g^{-1}(x)u(x,t)|^2=
|u(x,t)|^2$.

To prove the AB effect it is enough to show that  
for some $u(x,t)$ 
the probability density 
 $|u(x,t)|^2$  changes when we change 
the gauge equivalence class.

\section{The proof of the magnetic AB effect}
\label{section 3}
\init

\subsection{The case of one obstacle in $\R^2$}
Consider the Schr\"odinger equation (\ref{eq:1.1})   
in $(\R^2\setminus \Omega_1)\times(0,T)$  with the boundary condition (\ref{eq:1.2}) and the
initial condition (\ref{eq:1.3}).

Let 
$w(x,t)$  be the solution of the wave equation
\begin{equation}                                    \label{eq:3.1}
\frac{h^2}{2m}\frac{\partial^2 w}{\partial t^2}+H w=0\ \ \mbox{in}\ \ 
(\R^2\setminus \Omega_1)\times(0,+\infty)
\end{equation}
with the boundary condition
\begin{equation}                                       \label{eq:3.2}
w\big|_{\partial\Omega_1\times(0,+\infty)}=0
\end{equation}
and the initial conditions 
\begin{equation}                                      \label{eq:3.3}
w(x,0)=u_0(x),\ \  \frac{\partial w(x,0)}{\partial t}=0,\ \ x\in \R^2\setminus\Omega_1,
\end{equation}
i.e. $w(x,t)$ is even in $t$.
Here 
$$
H=\frac{1}{2m}\Big(-ih\frac{\partial}{\partial x}-\frac{e}{c}A\Big)^2+eV(x).
$$
There is a formula relating $u(x,t)$  and $w(x,t)$  (cf. Kannai [K]):
\begin{equation}                                        \label{eq:3.4}
u(x,t)=\frac{e^{-i\frac{\pi}{4}}\sqrt m}{\sqrt{2\pi ht}}
\int_{-\infty}^\infty e^{\frac{imx_0^2}{2ht}}w(x,x_0)dx_0.
\end{equation}
We shall consider  solutions of (\ref{eq:3.1}) such that
\begin{equation}                                       \label{eq:3.5}
|w(x,t)|\leq C(1+|t|)^m,\ \ 
\Big|\frac{\partial^r w(x,t)}{\partial t^r}\Big|
\leq C_r(1+|t|)^m,\ \ \forall r\geq 1.
\end{equation}
Let $\chi_0(t)\in C_0^\infty(\R^1),\ \chi_0(-t)=\chi_0(t),\ \chi_0(t)=1$
for  $|t|<\frac{1}{2},\ \chi_0(t)=0$   for  $|t|>1$.  We define  the integral
(\ref{eq:3.4})   as the limit of
\begin{equation}                                        \label{eq:3.6}
\frac{e^{-i\frac{\pi}{4}}\sqrt m}{\sqrt{2\pi ht}}
\int_{-\infty}^\infty \chi_0(\e x_0) e^{\frac{imx_0^2}{2ht}}w(x,x_0)dx_0
\end{equation}
as $\e\rightarrow 0$,  and we shall show that this limit exists  for any
$w(x,x_0)$ satisfying  (\ref{eq:3.5}).  Substitute the identity
$$
\Big(\frac{ht}{imx_0}\frac{\partial}{\partial x_0}\Big)^M
 e^{\frac{imx_0^2}{2ht}}=e^{\frac{imx_0^2}{2ht}},
\ \ \ \forall M,
$$
in (\ref{eq:3.6}) and integrate by parts in (\ref{eq:3.6}) for  $|x_0|>1$.
If $M\geq m+2$  we get an absolutely integrable function of $x_0$ and 
therefore we can pass to the limit when $\e\rightarrow 0$.

Note that
\begin{equation}                                    
\nonumber
\Big(-ih\frac{\partial}{\partial t}+H\Big)u(x,t)
=\frac{e^{-i\frac{\pi}{4}}\sqrt m}{\sqrt{2\pi ht}}
\int_{-\infty}^\infty e^{\frac{imx_0^2}{2ht}}
\Big(\frac{h^2}{2m}\frac{\partial^2}{\partial x_0^2}+H\Big)
w(x,x_0)dx_0,
\end{equation}
 Note also that
\begin{equation}
\nonumber
u(x,0)=\lim_{t\rightarrow 0}\frac{e^{-i\frac{\pi}{4}}\sqrt m}{\sqrt{2\pi ht}}
\int_{-\infty}^\infty e^{\frac{imx_0^2}{2ht}}w(x,x_0)dx_0=
w(x,0).
\end{equation}
Therefore $u(x,t)$  satisfies (\ref{eq:1.1}),  (\ref{eq:1.2}),  (\ref{eq:1.3})
if  $w(x,t)$  satisfies  (\ref{eq:3.1}),  (\ref{eq:3.2}),  (\ref{eq:3.3}).
\qed

We  shall  construct  geometric optics type solutions of
(\ref{eq:3.1}), (\ref{eq:3.2}), (\ref{eq:3.3}) and then use the formula 
(\ref{eq:3.4})  to obtain solutions of the Schr\"odinger equation.

We shall look for $w(x,t)$  in the form   
\begin{equation}                                \label{eq:3.7}
w_N(x,t)=e^{i\frac{mk}{h}(x\cdot\omega - t)}
\sum_{p=0}^N\frac{a_p(x,t)}{(ik)^p}+e^{i\frac{mk}{h}(x\cdot \omega+t)}
\sum_{p=0}^N\frac{b_p(x,t)}{(ik)^p},
\end{equation}
where $k$  is a large parameter  and $\omega$  is a unit vector,  i.e. $|\omega|=1$.

Substituting
(\ref{eq:3.7})  into (\ref{eq:3.1})  and equating equal  powers of $k$  we get
\begin{eqnarray}                                     \label{eq:3.8}
ha_{0t}(x,t)+h\omega\cdot a_{0x}(x,t)-i\omega\cdot \frac{e}{c}A(x)a_0=0,
\\
\nonumber
-hb_{0t}(x,t)+h\omega\cdot b_{0x}-i\omega\cdot \frac{e}{c} A(x)b_0=0,
\end{eqnarray}
\begin{eqnarray}                                     \label{eq:3.9}
\ \ 
ha_{pt}(x,t)+h\omega\cdot a_{px}(x,t)-i\omega\cdot \frac{e}{c}A(x)a_p=
i\Big(\frac{h^2}{2m}\frac{\partial^2}{\partial t^2}+H\Big)a_{p-1},
\\
\nonumber
-hb_{pt}(x,t)+h\omega\cdot b_{px}-i\omega\cdot \frac{e}{c} A(x)b_p=
i\Big(\frac{\partial^2}{\partial t^2}+H\Big)b_{p-1},\ \ 1\leq p\leq N.
\end{eqnarray}
We have $b_p(x,t)=a_p(x,-t)$  for $p\geq 0$,  assuming that $b_p(x,0)=a_p(x,0)$.

Introduce new coordinates $(s,\tau,t)$ instead  of $(x_1,x_2,t)$  where 
\begin{eqnarray}                                 \label{eq:3.10}
s=(x-x^{(0)})\cdot\omega -t,
\\
\nonumber
\tau=(x-x^{(0)})\cdot \omega_\perp,
\\
\nonumber
t=t.\ \ \ \ \ \ \ \ \ \ \ \ \ \ \ \ \ \  
\end{eqnarray}
Here
$\omega_\perp\cdot \omega=0,\ |\omega_\perp|=|\omega|=1$.  We assume that  
$x^{(0)}$  is a fixed point outside of the obstacle $\Omega_1$  
and that the line $x=x^{(0)}+s\omega,\ s\in \R,$  does not intersect $\Omega_1$.
Equations (\ref{eq:3.8}), (\ref{eq:3.9})  have the following form in the new
coordinates
\begin{eqnarray}                                         \label{eq:3.11}
\hat a_{0t}(s,\tau,t)
-i\omega\cdot\frac{e}{hc}A(x^{(0)}+(s+t)\omega+\tau\omega_\perp)\hat a_0=0,
\\
\nonumber
\hat a_{pt}(s,\tau,t)
-i\omega\cdot\frac{e}{hc}A(x^{(0)}+(s+t)\omega+\tau\omega_\perp)\hat a_p=
\hat f_p(s,\tau,t),\ \ p\geq 1,
\end{eqnarray}
where 
$\hat a_p(s,\tau,t)=a_p(x,t),\ \hat f_p(s,\tau,t)$ is 
$\frac{i}{h}\Big(\frac{h^2}{2m}\frac{\partial^2}{\partial t^2}+H\Big)a_{p-1}$
in the 
new coordinates.

We impose the following initial conditions
\begin{eqnarray}                                        \label{eq:3.12}
\hat a_0(s,\tau,0)=\frac{1}{2}\chi_0\Big(\frac{\tau}{\delta_1}\Big)
\chi_0\Big(\frac{s}{\delta_2 k}\Big),
\\
\nonumber
\hat a_p(s,\tau,0)=0\ \ \mbox{for}\ \ p\geq 1,
\end{eqnarray}
where 
$\chi_0(s)$  is the same as above.  We assume  that $\delta_1$  is  such that\linebreak  
$\mbox{supp\ }\chi_0(\frac{(x-x^{(0)})\cdot\omega_\perp}{\delta_1})$  does not 
intersect $\Omega_1$.
Then 
$$
\hat a_0(s,\tau,t)=\frac{1}{2}\chi_0\Big(\frac{\tau}{\delta_1}\Big)\chi_0
\Big(\frac{s}{\delta_2k}\Big)
\exp\Big(\frac{ie}{hc}\int_0^t\omega\cdot A(x^{(0)}+(s+t')
\omega+\tau\omega_\perp)dt'\Big).
$$
Since $s=(x-x^{(0)})\cdot \omega-t$  we have  in the original coordinates 
\begin{multline}                                         \label{eq:3.13}
a_0(x,t)=\frac{1}{2}\chi_0\Big(\frac{(x-x^{(0)})\cdot \omega_\perp}{\delta_1}\Big)
\chi_0\Big(\frac{(x-x^{(0)})\cdot \omega-t}{\delta_2k}\Big)
\\
\cdot
\exp\Big(\frac{i e}{hc}\int_0^t\omega\cdot A(x-t''w)dt''\Big),
\end{multline}
where we made the change of variables $t-t'=t''$.
Note  that
\begin{equation}                                    \label{eq:3.14}
|a_p(x,t)|\leq C t^p,\ \ 1\leq p\leq N,
\end{equation}
and (\ref{eq:3.5})  holds for any  $r\geq 1$.
Since  $b_p(x,t)=a_p(x,-t),\ p\geq 0,$   we
have that
\begin{eqnarray}                                         \label{eq:3.15}
\ \ \ \ \ \ \ 
w_N(x,0)=\chi_0\Big(\frac{(x-x^{(0)})\cdot \omega_\perp}{\delta_1}\Big)
\chi_0\Big(\frac{(x-x^{(0)})\cdot \omega}{\delta_2k}\Big)
e^{i\frac{m}{h}k\omega\cdot  x},
\\
\nonumber
w_{Nt}(x,0)=0.
\end{eqnarray}
Let 
\begin{equation}                                      \label{eq:3.16}
u_N(x,t)=\frac
{e^{-i\frac{\pi}{4}}\sqrt{m}}{\sqrt{2\pi ht}}\int_{-\infty}^\infty
e^{\frac{imx_0^2}{2ht}}w_N(x,x_0)dx_0.
\end{equation}
Using that $b_p(x,t)=a_p(x,-t)$ and making a change of variables we get
\begin{equation}                                       \label{eq:3.17}
u_N(x,t)=
\frac
{2e^{-i\frac{\pi}{4}}\sqrt{m}}{\sqrt{2\pi ht}}\int_{-\infty}^\infty
e^{\frac{imx_0^2}{2ht}+\frac{imk}{h}(x\cdot \omega-x_0)}
\sum_{p=0}^N\frac{a_p(x,x_0)}{(ik)^p}
dx_0.
\end{equation}
We have
\begin{equation}                                      \label{eq:3.18}
\Big(-ih\frac{\partial }{\partial t} +H\Big)u_N(x,t)=\frac
{e^{-i\frac{\pi}{4}}\sqrt{m}}{\sqrt{2\pi ht}}\int_{-\infty}^\infty
e^{\frac{imx_0^2}{2ht}}
\Big(\frac{h^2}{2m}\frac{\partial^2}{\partial x_0^2}+H\Big)
w_N(x,x_0)dx_0.
\end{equation}
Note that
\begin{multline}         
\Big(\frac{h^2}{2m}\frac{\partial^2}{\partial x_0^2}+H\Big)
w_N(x,x_0)=
e^{\frac{imk}{h}(x\cdot \omega-x_0)}
\Big(\frac{h^2}{2m}\frac{\partial^2}{\partial x_0^2}+H\Big)
a_N(x,x_0)
\\
\nonumber
+
e^{\frac{imk}{h}(x\cdot \omega+x_0)}
\Big(\frac{h^2}{2m}\frac{\partial^2}{\partial x_0^2}+H\Big)
b_N(x,x_0).
\end{multline}
Denote by $g_N(x,t)$ the  right hand  side  of (\ref{eq:3.18}).
Since $b_N(x,x_0)=a_N(x,-x_0)$  we have 
\begin{equation}                               \label{eq:3.19}
g_N(x,t)=
\frac 
{2e^{-i\frac{\pi}{4}}\sqrt{m}}{\sqrt{2\pi ht}}\int_{-\infty}^\infty
e^{\frac{imx_0^2}{2ht}}e^{\frac{imk}{h}(x\cdot\omega - x_0)}
\Big(\frac{h^2}{2m}\frac{\partial^2}{\partial x_0^2}+H\Big)
\frac{a_N(x,x_0)}{(ik)^N}dx_0.
\end{equation}
We apply 
the stationary phase method to the integral (\ref{eq:3.17}).
The equation for the critical point is 
$\frac{mx_0}{ht}-\frac{mk}{h}=0$,  i.e.  $x_0=kt$  and the Hessian is  
$\frac{m}{ht}$.  Therefore
\begin{multline}                                \label{eq;3.20}
u_N(x,t)=e^{-\frac{imk^2t}{2h}+i\frac{mk}{h}x\cdot\omega}
\chi_0\Big(\frac{(x-x^{(0)})\cdot\omega_\perp)}{\delta_1}\Big)
\exp\Big(\frac{ie}{hc}\int_0^\infty\omega\cdot A(x-s'\omega)ds'\Big)
\\
+O(\e),
\end{multline}
where
$\e$  is arbitrary small when $k$  is sufficiently large and
$t$  is sufficiently small.

We used that $\chi_0\big(\frac{x\cdot \omega -k t}{\delta_2 k}\big)=1$
when $k$  is large and $t$  is small.
Note that 
\begin{equation}                              \label{eq:3.21}
\Big|\frac{a_p(x,kt)}{(ik)^p}\Big|
\leq \frac{C}{k^p}(kt)^p\leq Ct^p
\end{equation}is small when $t$  is small.

Applying the stationary phase method to (\ref{eq:3.19}) we get,  using (\ref{eq:3.14}),
that 
\begin{equation}                                \label{eq:3.22}
\int_{\R^2\setminus\Omega}|g_N(x,t)|^2dx\leq C t^{2N}k.
\end{equation}We used in (\ref{eq:3.22})  that 
$\chi_0\big(\frac{x\cdot \omega -k t}{\delta_2 k}\big)=0$
when  $|x\cdot \omega|>Ck$.

Denote  by  $\|g_N\|_r$  the Sobolev norm in $H_r(\R^2\setminus\Omega_1)$.
It follows from (\ref{eq:3.19})  and (\ref{eq:3.14})  that 
$$
\|g_N\|_r\leq Ct^Nk^{r+\frac{1}{2}}.
$$
Let $R_N(x,t)$  be the solution of 
$$
\Big(-ih\frac{\partial}{\partial t}+H\Big)R_N=-g_N(x,t)
\ \ \mbox{in} \ \ (\R^2\setminus\Omega_1)\times(0,T),
$$
$$
R_N\Big|_{\partial\Omega_1\times(0,T)}=0,
$$
$$
R_N(x,0)=0.
$$
Such solution exists and satisfies the following estimates (cf. [E1]):
\begin{equation}                                   \label{eq:3.23}
\max_{0\leq t\leq T}\|R_N(\cdot,t)\|_2
\leq C\int_0^T\Big( \|g_N(\cdot,t)\|_0
+\big\|\frac{\partial g_N(\cdot,t)}{\partial t}\big\|_0\Big)dt.
\end{equation}
By the Sobolev embedding theorem
 $|R_N(x,t)|\leq C\max_{0\leq t\leq T}\|R_N(\cdot,t)\|_2$ for all
$(x,t)\in (\R^2\setminus\Omega_1)\times(0,T)$.  Since
$$
\max_{0\leq t\leq T} \Big\|\frac{\partial^p}{\partial t^p}g_N(x,t)\Big\|_r\leq 
CT^Nk^{\frac{1}{2}+r}
$$
we get that 
$$
|R_N(x,t)|\leq C\e
$$
if $T\leq \frac{C}{k^{\delta_3}},\  0<\delta_3<1, \ (N+1)\delta_3>\frac{1}{2},\ 
k$ is large.

Note that 
$u=u_N+R_N$  satisfies (\ref{eq:1.1}), (\ref{eq:1.2}) and the initial condition
$u(x,0)=u_N(x,0)=\chi_0(\frac{(x-x^{(0)})\cdot\omega_\perp}{\delta_1})
\chi_0(\frac{(x-x^{(0)})\cdot\omega}{\delta_2k})e^{i\frac{mk}{h}x\cdot \omega}$.
  Therefore we constructed a solution $u(x,t,\omega)$  for
$x\in \R^2\setminus\Omega_1,\ t\in (0,T),\ T=O(\frac{1}{k^\delta_3}),\ k$
is large,  such that
\begin{multline}                                       \label{eq:3.24}
u(x,t,\omega)
\\
=e^{-i\frac{mk^2t}{2h}-i\frac{mk}{h}x\cdot\omega}
\chi_0\Big(\frac{(x-x^{(0)})\cdot\omega_\perp}{\delta_1}\Big)
\exp\Big(i\frac{e}{hc}\int_0^\infty \omega\cdot A(x-s'\omega)ds'\Big)
+O(\e),
\end{multline}
where $\e$  can be chosen arbitrary small  if  $k$  is large enough.
Note that the integral in (\ref{eq:3.24}) converges since $A(x)=C\frac{(x_2,-x_1)}{|x|^2}
+O(\frac{1}{|x|^2}).$
\qed

Let $x^{(0)}\in \R^2\setminus\Omega_1$  and let $\omega$  and $\theta$  be
two unit vectors (see Fig.1):

\begin{tikzpicture}

\draw[->](-5,0)--(6,0);
\draw[->](0,-2.2)--(0,9.2);

\draw[->](-5,-2)--(0.25,8.5); 
\draw(0.5,8.5) node {$\theta$};
\draw[->](5,-2)--(-0.25,8.5); 
\draw(-0.5,8.5) node {$\omega$};

\draw(-5.5,-1.5)--(5.5,-1.5);

\draw(0,8)--(4,0);
\draw(0,8)--(-4,0);

\draw (0.5,3) circle (1.2 cm);
\draw (0.5,3.2) node {$\Omega$};
\draw(1.5,8) node {$x^{(0)}=(0,L)$};

\draw(0.2,-0.3)  node {$0$};
\draw(4.8,-2.3) node {$\small{(N\sin\varphi,-N\cos\varphi +L)}$};
\draw(-4.6,-2.3) node {$(-N\sin\varphi,-N\cos\varphi +L)$};


\draw(6,-0.3) node  {$x_1$};
\draw(0.4,9.1) node {$x_2$};

\draw(0,6) arc (270:325:1 and 0.8);
\draw (0.4,6.5) node {$\varphi$};

\fill (-4.75,-1.5) circle (2pt);
\fill (4.75,-1.5) circle (2pt);
\fill (0,8) circle (2pt);

\end{tikzpicture}

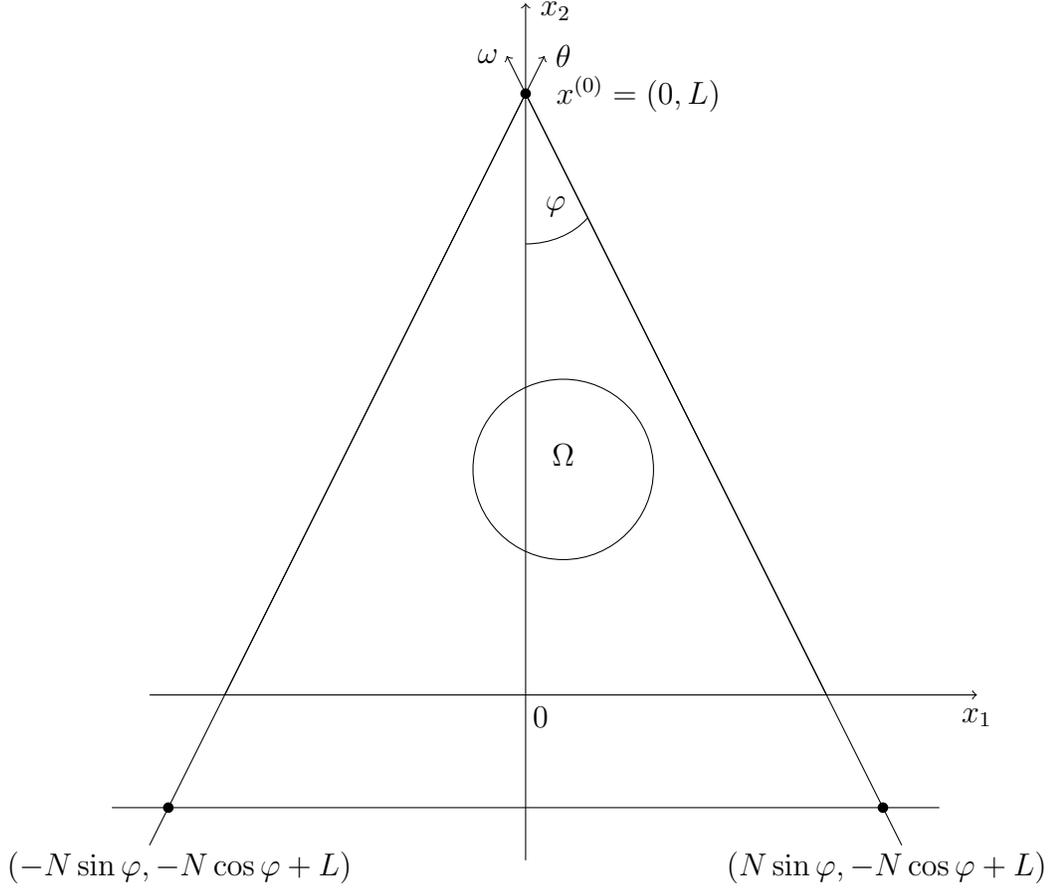
 Figure 1.
\\
\\

Consider the difference 
of two solutions of the form (\ref{eq:3.24}) corresponding to
$(x^{(0)},\omega)$  and  $(x^{(0)},\theta)$,  respectively:
\begin{equation}                               \label{eq:3.25}
w(x,t)=u(x,t,\omega)-u(x,t,\theta),
\end{equation}
where $u(x,t,\omega)$ is the same as in (\ref{eq:3.24}) and
\begin{multline}                                       \label{eq:3.26}
u(x,t,\theta)
\\
=e^{-i\frac{mk^2t}{2h}+\frac{imk}{h}x\cdot\theta}
\chi_0\Big(\frac{(x-x^{(0)})\cdot\theta_\perp}{\delta_1}\Big)
\exp\Big(i\frac{e}{hc}\int_0^\infty \theta\cdot A(x-s'\theta)ds'\Big)+O(\e),
\end{multline}
where $\theta_\perp\cdot\theta=0$.  Note that
modulo $O(\e)$  the support of $v_1$  is contained  in a small neighborhood 
of the line $x=x^{(0)}+s\omega$  and the support of $v_2$  is contained in
a small neighborhood of $x=x^{(0)}+s\theta$.

Let $U_0$  be a disk of radius $\e_0$ centered in $x^{(0)}$ and  contained in
$(\mbox{supp\ }v_1)\cap(\mbox{supp\ }v_2)$.
We assume that $\chi_0(\frac{(x-x^{(0)})\cdot\omega_\perp}{\delta_1})=
\chi_0(\frac{(x-x^{(0)})\cdot\theta_\perp}{\delta_1})=1$  
in $U_0$. We have for $x\in U_0$
and $0<t<T=\frac{1}{k^{\delta_3}}$
\begin{multline}                                              \label{eq:3.27}
|u(x,t,\omega)-u(x,t,\theta)|^2=
\Big|1-e^{i\frac{mk}{h}x\cdot(\omega-\theta)+i(I_1-I_2)}\Big|^2+O(\e)
\\
=4\sin^2\frac{1}{2}\Big(\frac{mk}{h}x\cdot(\omega-\theta)+I_1-I_2\Big)+O(\e),
\\
I_1=\frac{e}{hc}\int_0^\infty\omega\cdot A(x-s\omega)ds,\ \ \
I_2=\frac{e}{hc}\int_0^\infty\theta\cdot A(x-s\theta)ds,
\end{multline}    
and $k>k_0,\ k_0$  is large,  $T\leq \frac{1}{k_0^{\delta_3}}$.

Choose $k_n>k_0$  such  that
\begin{equation}                                       \label{eq:3.28}
\frac{mk_n}{h}x^{(0)}\cdot(\omega-\theta)=2\pi n,\ \ n\in\Z.
\end{equation}
Let, for simplicity,  $\theta_1=-\omega_1,\ \theta_2=\omega_2,\ x^{(0)}=(0,L),\ 
\tan\varphi=\frac{\theta_1}{\theta_2}$  is small.  Define 
$$
I_{1N}(x,\omega)=
\frac{e}{hc}\int_0^{N}\omega\cdot A(x-s\omega)ds,
$$
$$
I_{2N}(x,\theta)=\frac{e}{hc}\int_0^N\theta\cdot A(x-s\theta)ds,
$$
$$
I_{3N}=\frac{e}{hc}\int_{-N\sin\varphi}^{N\sin\varphi}A_1(s,-N\cos\varphi+L)ds
$$
(see Fig.1).
Note 
that 
\begin{equation}                                          \label{eq:3.29}
-I_{1N}(x^{(0)},\omega)+I_{2N}(x^{(0)},\theta)+I_{3N}=\alpha,
\end{equation}
where $\alpha$ is the magnetic flux  (cf. (\ref{eq:1.4})).   We assume that
\begin{equation}                                        \label{eq:3.30}
\alpha\neq 2\pi n,\ \forall n\in \Z.
\end{equation} 
Since $|A|\leq\frac{C}{r}$,  where  $r$
is the distance to $\Omega_1$,
we have
\begin{equation}                                       \label{eq:3.31}
|I_{3N}|\leq\frac{e}{hc}\frac{C}{N}2N\sin\varphi=
C_1\frac{e}{hc}\sin\varphi.
\end{equation}
When $N\rightarrow\infty$  we get
\begin{equation}                                        \label{eq:3.32}
I_2-I_1=\alpha+O\Big(\frac{e}{hc}\sin\varphi\Big)\ \ \mbox{for}\ \ x\in U_0.
\end{equation}
Assuming that the radius of the disk $U_0$  is $\e_0$  we get
\begin{equation}                                      \label{eq:3.33}
\Big|\frac{mk_n}{h}(x-x^{(0)})\cdot(\omega-\theta)\Big|\leq 2\frac{mk_n}{h}\e_0\sin\varphi.
\end{equation}
Therefore
for arbitrary small $\e>0$
 using (\ref{eq:3.28}), (\ref{eq:3.29}),  (\ref{eq:3.31}),  (\ref{eq:3.32}),  (\ref{eq:3.33}),
fixing  $k_n>k_0$  and  choosing  $\varphi$  and  $\e_0$  small enough we get
\begin{equation}                                      \label{eq:3.34}
|u(x,t,\omega)-u(x,t,\theta)|^2=4\sin^2\frac{\alpha}{2}+O(\e),
\end{equation}
where $x\in U_0$  and $\alpha$  is the magnetic flux (\ref{eq:1.4}).

Therefore,  Theorem \ref{theo:1.1} is proven.

\subsection{The three-dimensional case}

The constructions of the subsection 3.1 can be carried out  in the case of three 
dimensions.  Consider,  for example, a toroid  $\Omega_1$  in $\R^3$  as in 
Tonomura et al experiment (cf [T et al]).  Let $x^{(0)}$  be a point outside
 of $\Omega_1$
and let $\gamma_1=\{x=x^{(0)}+s\omega,\ s\leq 0\}$
be a ray passing  through the hole of the toroid.  As in subsection 3.1  we can construct a solution $v_1(x,t,\omega)$  of the form (\ref{eq:3.26}).
In the case $n\geq 3$  dimensions there are $(n-1)$  orthogonal unit vectors
$\omega_{\perp j},1\leq j\leq n-1$,  such that $\omega\cdot\omega_{\perp j}=0,\ 
1\leq j\leq n-1$,  and  we have to replace 
$\chi_0(\frac{(x-x^{(0)})\cdot\omega_\perp}{\delta_1})$
in (\ref{eq:3.26})  by
$\Pi_{j=1}^{n-1}\chi_0(\frac{(x-x^{(0)})\cdot\omega_{\perp j}}{\delta_1})$.
Let
$\gamma_2=\{x=x^{(0)}+s\theta,s\leq 0\}$
be a ray  passing  outside  of toroid and
 let  $v_2(x,t,\omega)$  be the corresponding solution  
of the form (\ref{eq:3.27}).  As in subsection 3.1  we get
$$
|v_1(x,t,\omega)-v_2(x,t,\theta)|^2=4\sin^2\frac{\alpha}{2}+O(\e),
$$
where $\alpha=\frac{e}{hc}\int_\gamma A(x)\cdot dx,\ \gamma$  is a closed simple curve
encircling $\Omega_1$  and we assume  that  the angle between $\omega$  and $\theta$
is small.  

Assuming that $\alpha\neq 2\pi n,\forall n\in \Z$,  we obtain that the 
probability density $|v_1-v_2|^2$  depends on $\alpha$  and this proves 
the AB effect.

\subsection{The case of several  obstacles}

Let $\Omega_j,\ 1\leq j\leq m,\ m>1,$
be  obstacles in $\R^2$,  and let  $\alpha_j=\frac{e}{hc}\int_{\gamma_j} A(x)\cdot dx$ 
be the magnetic fluxes generated by magnetic fields shielded in $\Omega_j,\ 
1\leq j\leq m$.  Suppose that some $\alpha_j$  satisfy  the condition (\ref{eq:3.30}).
If the obstacles are close to each other  it 
is impossible to repeat the construction of the subsection 3.1 separately 
for each  $\Omega_j$.  Note  that if the total flux 
$\sum_{j=1}^m\alpha_j=2\pi p,\ p\in\Z$,  then  treating $\Omega=\cup_{j=1}^m\Omega_j$ 
as one obstacle  we will miss the magnetic AB effect.

We shall  introduce some notations.

Let  $x^{(1)}\not\in \Omega=\cup_{j=1}^m\Omega_j$.  Denote by 
$\gamma=\gamma_1\cup\gamma_2\cup...\cup\gamma_r$  the broken ray starting 
at $x^{(1)}$  and reflecting at $\Omega$  at points $x^{(2)},...,x^{(r)}$.
The last  leg $\gamma_r$  can be extended to the infinity.
Denote  by $\omega_p,\ 1\leq p\leq r$,  the directions of $\gamma_p$.
The equations of $\gamma_1,...,\gamma_r$  are $x=x^{(1)}+s\omega_1,\ 
s_1=0\leq s\leq s_2,\ x=x^{(2)}+s\omega_2,
\ s_2\leq s\leq s_3,...,x=x^{(r)}+s\omega_r,\ s_r\leq s<+\infty$.
Here $s_p$  are  such that $x(s_p)=x^{(p)},\ 1\leq p\leq r$.
Denote by 
$\tilde\gamma=\tilde\gamma_1\cup\tilde\gamma_2\cup...\cup\tilde\gamma_r$ 
the lifting of $\gamma$  to $\R^2\times(0,+\infty)$,  where
the equations of $\tilde\gamma_p$  are  
$x=x^{(p)}+s\omega_p,\ t=s,\ s_p\leq s\leq s_{p+1},\ s_{r+1}=+\infty$.
Note  that the times when $\tilde\gamma$  hits the obstacles are 
$t_p=s_p,\ 2\leq p\leq r$.

Let $V_0$ be a small neighborhood 
 of $x^{(1)}$.  Denote by  $\gamma_y=\cup_{p=1}^r\gamma_{py}$  the broken ray that starts
at  $y\in V_0$.
 We assume  that $\gamma_{1y}$  has  the form $x=y+s\omega_1,\ 0\leq s\leq s_2(y),$
where  $x^{(2)}(y)=y+s_2(y)\omega_1$  is the point where
$\gamma_{1y}$  hits  $\partial\Omega$.  We have
$\gamma_{x^{(1)}}=\gamma$.  Let  $U_0(t)=\{x=x(t)\}$  be the set  of endpoints 
at the time $t$  of $\tilde\gamma_{ry},\ y\in V_0$.  Note that there is a
one-to-one correspondence between $y\in V_0$  and $x(t)\in U_0(t)$.
Therefore  we shall denote the broken ray starting  at $y\in V_0$  and
ending at  $x(t)$  at the time $t$ by  $\gamma(x(t))$  instead of $\gamma_y$.
As in  [E6]  we can construct a geometric optics solution of
$(\frac{h^2}{2m}\frac{\partial^2}{\partial t^2}+H)w_N=0$  in
$(\R^2\setminus\Omega)\times(0,+\infty)$  in the form
\begin{multline}                                   \label{eq:3.35}
w_N(x,t)
\\
=\sum_{p=1}^r\sum_{n=0}^Ne^{i\frac{mk}{h}(\psi_p(x)-t)}\frac{a_{pn}(x,t)}{(ik)^n}
+
\sum_{p=1}^r\sum_{n=0}^Ne^{i\frac{mk}{h}(\psi_p(x)+t)}\frac{b_{pn}(x,t)}{(ik)^n},
\end{multline}
where
\begin{align}                                     \label{eq:3.36}
|\nabla \psi_p(x)|=1,
&\ \ \frac{\partial\psi_p(x^{(p)})}{\partial x}=\omega_p,\ \ 1\leq p\leq r,
\\
\nonumber
&
\psi_1(x)=x\cdot\omega_1.
\end{align}
We have that $a_{pn}(x,t)=b_{pn}(x,-t)$
and $a_{pn}(x,t)$ satisfy the transport equations
\begin{multline}                                \label{eq:3.37}
\frac{\partial a_{pn}}{\partial t}+\frac{\partial\psi_p(x)}{\partial x}
\cdot\frac{\partial a_{pn}}{\partial x}+\frac{1}{2}\Delta\psi_p a_{pn}
-i\frac{e}{hc}A(x)\cdot\frac{\partial\psi_p}{\partial x}a_{pn}=f_{pn}(x,t),
\\
1\leq p\leq r,\ 0\leq n\leq N,\qquad\qquad
\end{multline}
where
$f_{p0}=0,\ f_{pn}$  depend on $a_{pj}$  for $n\geq 1,\ 0\leq j\leq n-1$.
The following boundary conditions hold on $\partial\Omega\times(0,+\infty)$:
\begin{align}                                  \label{eq:3.38}
&\psi_p\big|_{\partial\Omega\times(0,+\infty)}
=\psi_{p+1}\big|_{\partial\Omega\times(0,+\infty)},\ \ 1\leq p\leq r-1,
\\
\nonumber
&a_{pn}\big|_{\partial\Omega\times(0,+\infty)}=
-a_{p+1,n}\big|_{\partial\Omega\times(0,+\infty)},\ \  1\leq p\leq r-1.
\end{align}
Conditions (\ref{eq:3.38}) imply  that
$$
w_N\Big|_{\partial\Omega\times(0,+\infty)}=0.
$$
We  impose the following initial conditions:
\begin{align}                                 \label{eq:3.39}
&a_{10}(x,0)=\frac{1}{2}\chi_0\Big(\frac{(x-x^{(1)})\cdot\omega_{1\perp}}{\delta_1}\Big)
\chi_0\Big(\frac{(x-x^{(1)})\cdot\omega_1}{\delta_2 k}\Big),
\\
\nonumber
&a_{1n}(x,0)=0,\ \ n\geq 1.
\end{align}
We assume that 
$\delta_1,\delta_2$  in (\ref{eq:3.37}) are small,  so that the support of
 the first sum  in (\ref{eq:3.35})  is contained in a small 
neighborhood
of $\tilde\gamma=\cup_{p=1}^r\tilde\gamma_p$.  
We define $a_{pn}(x,t)$  as zero  outside of this neighborhood of $\tilde\gamma$.

Let $x^{(0)}\in\gamma_r$  and  $(x^{(0)},t^{(0)})$ be a corresponding  point  on
$\tilde\gamma_r$.
It was shown in  [E6] that
\begin{equation}                                    \label{eq:3.40}
a_{r0}(x,t)=c_0(x,t)\exp\Big(\frac{ie}{hc}\int_{\tilde\gamma(x,t)}A(x)\cdot dx\Big) 
+O\Big(\frac{1}{k}\Big),
\end{equation}
where
$\tilde\gamma(x,t)$  is the broken  ray  starting  in a neighborhood
of $x^{(1)}$  at $t=0$  and ending  at $(x,t),\ c(x,t)\neq 0$
in a neighborhood of $(x^{(0)},t^{(0)})$. 

As in subsection 3.1  we have that $a_{rn}(x,t),n\geq 1$
satisfy  the estimates  of the form  (\ref{eq:3.5}).

Let  $\tilde\gamma(x^{(0)},t^{(0)})$  be  the broken ray starting  at
$(x^{(1)},0)$  and ending  at $(x^{(0)},t^{(0)})$,  where  $x^{(0)}\in \gamma_r$.
Let
\begin{equation}                                        \label{eq:3.41}
u_N(x,t)=\frac{e^{-i\frac{\pi}{4}}\sqrt m}{\sqrt{2\pi ht}}
\int_{-\infty}^\infty e^{\frac{imx_0^2}{2ht}}w_N(x,x_0)dx_0
\end{equation}
where $w_N(x,x_0)$  is  the same as in (\ref{eq:3.35}).
We assume in this subsection  that  
\begin{equation}                                      \label{eq:3.42}
t=\frac{t'}{k},\ \ 0\leq t'\leq T'. 
 \end{equation}
Applying the stationary phase method  to (\ref{eq:3.41})
and using (\ref{eq:3.40}), (\ref{eq:3.42})  we get for $x$  belonging to
a neighborhood of $x^{(0)}$
\begin{multline}                                    \label{eq:3.43}
u_{N}(x,t)
\\ 
=\exp\Big(i\big(-\frac{mk^2t}{2h}+\frac{mk}{h}\psi_r(x)\big)\Big)
c_0(x,t')\exp\Big(\frac{ie}{hc}\int_{\tilde\gamma(x,t')}A(x)\cdot dx\Big)
\\
 +O\Big(\frac{1}{k}\Big),
\end{multline}
where $t'=kt,\ t'$  belongs to a neighborhood of $t^{(0)}$.

Since $A(x)$  is independent of $t$  we have 
$$
\int_{\tilde\gamma(x,t')}A(x)\cdot dx=\int_{\gamma(x(t'))}A(x)\cdot dx,
$$
where $\gamma(x(t'))$  is the projection of $\tilde\gamma(x,t')$  on
the $x$-plane.

Analogously to subsection 3.1 we get  that there exists $R_N(x,t)$
such that $R_N(x,t)=O(\frac{1}{k}),\ t=\frac{t'}{k},\ 0\leq t'\leq T'$,  and
\begin{equation}                                   \label{eq:3.44}
u(x,t)=u_N(x,t)+R_N(x,t)
\end{equation}
is the exact solution of (\ref{eq:1.1})  with the boundary
conditions $u\big|_{\partial\Omega\times(0,\frac{T'}{k})}=0$  and the initial
condition
\begin{equation}                                 \label{eq:3.45}
u(x,0)=\chi_0\Big(\frac{(x-x^{(1)})\cdot\omega_{1\perp}}{\delta_1}\Big)
\chi_0\Big(\frac{(x-x^{(1)})\cdot\omega_1}{\delta_2k}\Big)e^{i\frac{mk}{h} x\cdot \omega}.
\end{equation}
\qed

Denote by $\tilde\beta=\{x=x^{(2)}+s\theta,\ t=s,\ 0\leq s\leq t^{(0)}\}$
the ray
starting at $(x^{(2)},0)$  and  ending  exactly at  
the point $(x^{(0)},t^{(0)})$  (see Fig. 2).
Analogously to (\ref{eq:3.27})  we can construct a solution $v(x,t)$
of (\ref{eq:1.1}),  satisfying  (\ref{eq:1.2})
and such  that
\begin{multline}                             \label{eq:3.46}
v(x,t)=\chi_0\Big(\frac{(x-x^{(2)})\cdot\theta_{\perp}}{\delta_1}\Big)
\chi_0\Big(\frac{(x-x^{(2)})\cdot\theta-kt}{\delta_2 k}\Big)c_1(x,t')
\\
\cdot
\exp\Big(-i\frac{mk^2t}{2h}+i\frac{mk}{h}x\cdot\theta
+\frac{ie}{hc}\int_{\tilde\beta(x,t')}A(x)\cdot dx\Big)
+O\Big(\frac{1}{k}\Big),
\end{multline}
where 
$t=\frac{t'}{k},\ (x,t')\in U_0$,  where $U_0$  is a neighborhood of
 $(x^{(0)},t^{(0)})$.

We choose initial conditions $c_1(x,0)$ such that  (cf.  (3.43))
$$
c_1(x^{(0)},t^{(0)})=c(x^{(0)},t^{(0)}).
$$
Note that 
$$
\int_{\tilde\beta(x,t')}A\cdot dx=\int_{\beta(x(t'))}A(x)\cdot dx,
$$
where $\beta(x(t'))$  is the projection of $\tilde\beta(x,t')$.

As in (\ref{eq:3.28}) we have for $(x,t)$ near $(x^{(0)},\frac{t^{(0)}}{k})$
$$
|u(x,t)-v(x,t)|^2=|c(x^{(0)},t^{(0)})|^24\sin^2\frac{1}{2}
\Big(\frac{mk}{h}(\psi_r(x)-\theta\cdot x)+I_1-I_2\Big)+O(\e),
$$
where 
$$
I_1=\frac{e}{hc}\int_{\gamma(x^{(0)})} A\cdot dx,\ \ \ \
I_2=\frac{e}{hc}\int_{\beta(x^{(0)})} A\cdot dx.
$$
Choose  $k_n>k_0$ such  that
$$
\frac{mk_n}{h}(\psi_r(x^{(0)})-x^{(0)}\cdot\theta)=2\pi n,\ n\in \Z,
$$
and choose the initial points $x^{(1)}$ on $\gamma_1$  and $x^{(2)}$  on $\beta$
far enough from $\Omega$ to have the integral 
$$
I_3=\frac{e}{hc}\int_\sigma A\cdot dx
$$
small.  Here $\sigma$  is the straight  line connecting   $x^{(1)}$  and  $x^{(2)}$
and not  intersecting $\Omega$  (see Fig. 2).  Then if the neighborhood $U_0$
is small enough we get
\begin{equation}                                    \label{eq:3.47}
|u(x,t)-v(x,t)|^2=
|c(x^{(0)},t^{(0)})|4\sin^2\frac{\alpha}{2}+O(\e),
\end{equation}
where  $\alpha=I_1-I_2+I_3$  at $(x^{(0)},t^{(0)})$,
$(x,t)$ is close  to $(x^{(0)},\frac{t^{(0)}}{k})$.
Note that $\alpha$  is the sum of magnetic fluxes of obstacles that are bounded  
by  $\gamma\cup (-\beta)\cup\sigma$.  

We shall assume that this $\alpha\neq 0\ (\mbox{modulo\ } p,\ p\in Z)$.
Analogously to Theorem \ref{theo:1.1} we get from (\ref{eq:3.47}) the following result.
\begin{theorem}                                      \label{theo:3.1}
There exists a solution $u(x,t)$  of (\ref{eq:1.1}), (\ref{eq:1.2}) with highly 
oscillating initial data such that (\ref{eq:3.47})  holds in a neighborhood of 
$x^{(0)}=x(t^{(0)})$.  It follows from (\ref{eq:3.47})
that the magnetic potential affects the probability density $|u(x,t)|^2$  near $x^{(0)}=x(t^{(0)})$:
 the change in $\alpha$ leads to the change in $|u(x,t)|^2$.
This proves the Aharonov-Bohm effect in the case of several obstacles. 
\end{theorem}
Varying appropriately the broken rays we can recover all 
$\alpha_j(\mbox{mod\,} 2\pi p),\linebreak 1\leq j\leq m,$ 
up to a sign. 
\\
\\
\begin{tikzpicture}
\
\\
\\
\draw[->](0,0)--(2,2);
\draw[->](2,2)--(5,3);
\draw[->](5,3)--(3,7);
\draw[->](3,7)--(-1,8);
\draw[->](0,0)--(-1,8);

\draw (6,2.5) circle (1.12);
\draw (4,7.5) circle (1.12);
\draw (2.5,4.5) circle (1.2);

\draw (0,-0.5) node {$x^{(2)}=x(0)$};
\draw (3,1.7) node {$x^{(1)}=x(0)$};
\draw (4.2,2.4) node {$\gamma_1$};
\draw (1,0.7) node {$\sigma$};
\draw (4.4,5) node {$\gamma_2$};
\draw (2,7.5) node {$\gamma_3$};
\draw (-1,8.5) node {$x^{(0)}=x(t^{(0)})$};
\draw (-0.8,4) node {$\beta$};

\draw (6,2.5) node {$\Omega_1$};
\draw (4,7.5) node {$\Omega_2$};
\draw (2.5,4.5) node {$\Omega_3$};

\end{tikzpicture}

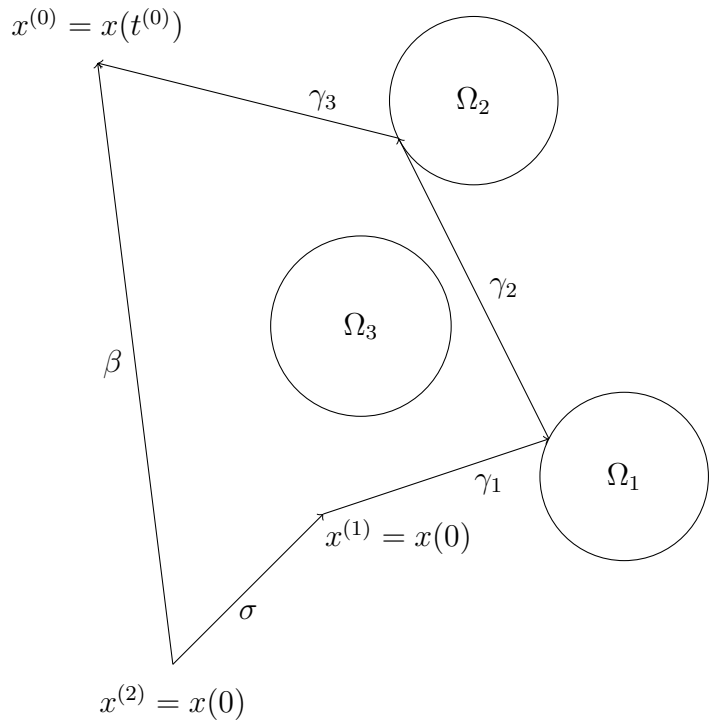
 Figure 2.
\\
\\

{\bf Remark 3.1.}
The broken rays are useful  also in the case of one obstacle.
 Consider two rays $\gamma_1$  and $\gamma_2$  
starting at the same point $P_0$  reflecting  from the  artificial  boundaries 
(mirrors)  $M_1$  and $M_2$  and merging  at some point $P_1$  behind the obstacle 
$\Omega$ (see Fig. 3). 
\\
\\
\begin{tikzpicture}

\draw (-6,0)--(5,0);
\draw (0,0)--(3,2);
\draw (3,2)--(-2.8,6);
\draw (-4,4)--(-2.8,6);
\draw (-4,4)--(0,0);
\draw (-4,3.5)--(-4,4.5);
\draw (3,1.5)--(3,2.5);

\draw (-4.5,4) node {$M_2$};
\draw (3.5,2) node {$M_1$};
\draw (0,-0.5) node {$P_0$};
\draw (-3,6.2) node {$P_1$};

\draw (-0.5,2.5) circle (1);
\draw (-0.5,2.5) node {$\Omega$};
\draw (1.5, 0.7) node {$\gamma_1$};
\draw (0,4.5) node {$\gamma_1'$};
\draw (-3,2.5) node {$\gamma_2$};
\draw (-3.5,5.5) node {$\gamma_2'$};

\end{tikzpicture}

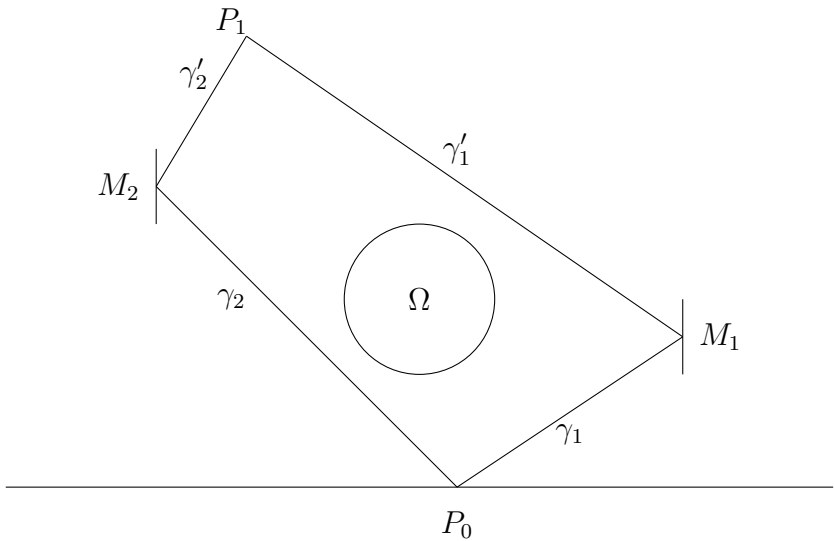
 Figure 3.
\\
\\
  One can construct solutions corresponding to the broken rays 
$\gamma_1\cup\gamma_1'$ and $\gamma_2\cup\gamma_2'$  and prove  the magnetic
AB effect.  Note  that this proof mimics the classical AB experiment where the beam
of electrons is splitted  at $P_0$ into two beams.   
Each of these beams reflects at $M_1$  and $M_2$  respectively  and they merge  
at the interferometer at $P_1$.

In [BW1], [BW2]  Ballesteros and Weder  study  the mathematical justification 
of the Tonomara and al. experiment by considering  high frequency  solutions
of the Schr\"odinger equations concentrating along one straight ray.
The use of broken rays may help to complete their results by 
taking into account 
  the
 splitting, reflecting and merging of rays.

\section{The proof  of the electric Aharonov-Bohm effect}
\label{section 4}
\init

Let $D$  be a domain in $\R^n\times[0,T]$ and let $D_{t_0}=
D\cap\{t=t_0\}$.  Assume that $D_{t_0}$  depends continuously on
$t_0\in [0,T]$  and that normals to $D\setminus(\overline{D_0}\cup
\overline{D_T})$
 are not parallel to the $t$-axis.
Suppose that  the magnetic potential $A(x,t)=0$  in $D$  and
consider the Schr\"odinger equation:
\begin{equation}                                     \label{eq:4.1}
ih\frac{\partial u(x,t)}{\partial t}+\frac{h^2}{2m}\Delta u(x,t)- e V(x,t)u(x,t)=0,\ \ 
(x,t)\in D,
\end{equation}             
with zero Dirichlet boundary condition
\begin{equation}                             \label{eq:4.2}
u\big|_{\partial D_t}=0\ \ \mbox{for}\ \ 0<t<T
\end{equation}
and nonzero initial condition
\begin{equation}                             \label{eq:4.3}
u(x,0)=u_0(x),\ \ x\in D_0.
\end{equation}
Suppose that electric field $E=\frac{\partial V}{\partial x}=0$  in $D$.
If $D_t$  are connected for all  $t\in [0,T]$  then 
$V(x,t)=V(t),$  i.e.  $V(t)$  does not depend  on $x$.  Making  the gauge transformation
\begin{equation}                                \label{eq:4.4}
v(x,t)=\exp\Big(i\frac{e}{h}\int_0^t V(t')dt'\Big)u(x,t)
\end{equation}
we get  that $v(x,t)$  satisfies the Schr\"odinger equation
\begin{equation}                                 \label{eq:4.5}
ih\frac{\partial v}{\partial t}+\frac{h^2}{2m}\Delta v(x,t)=0,
\end{equation}
where 
\begin{equation}                                  \label{eq:4.6}
v\big|_{\partial D_t}=0\ \ \mbox{for}\ \ 0<t<T,
\end{equation}
\begin{equation}                                  \label{eq:4.7}
v(x,0)=u_0(x),\ x\in D_0.
\end{equation}Therefore $V(t)$  is gauge equivalent  to zero electric potential, 
 i.e. there is no electric AB effect in the case when $D_{t_0}$  are connected 
for all $t_0\in [0,T]$.  To have the electric AB effect the domain $D$ 
 must be not connected on some subintervals of $(0,T)$.
\qed

We shall describe an electric AB  effect when $A=0, E=0$  
in $D$  but the electric potential $V(x,t)$  is not gauge equivalent to the zero potential.

Denote  by $\Omega(\tau)$  the interior  of 
the unit disk  $x_1^2+x_2^2\leq 1$ 
with removed 
 two  
 parts $\Delta(\tau)$  and  $\Delta(-\tau)$  depending on the parameter
$\tau,\ 0\leq \tau\leq \frac{1}{2}$
 (see Fig. 4).

\begin{tikzpicture}
\draw [->](-2.5,0)--(2.5,0);
\draw [->](0,-2.5)--(0,2.5);
\draw (0,0) circle (2);
\draw (-1.9,0.5)--(-0.5,0.5);
\draw (-1.9,-0.5)--(-0.5,-0.5);
\draw (-0.5,-0.5)--(-0.5,0.5);
\draw (0.5,-0.5)--(1.9,-0.5);
\draw (0.5,0.5)--(1.9,0.5);
\draw (0.5,-0.5)--(0.5,0.5);
\draw (-0.8,0.2) node {$-\tau$};
\draw (0.8,0.2)  node {$\tau$};
\draw (2.8,0)  node {$x_1$};
\draw (0,2.7) node {$x_2$};

\end{tikzpicture}
\\
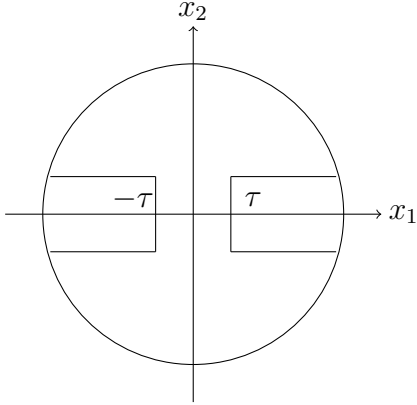
Figure 4.
\\
\\
Let $D$  be  the following domain in $\R^2\times[0,T+1]$:
$$
D_t=\Omega(\frac{1}{2}-t)\ \ \mbox{for}\ \ 0\leq t\leq \frac{1}{2},
$$
$$
D_t=\Omega(0)\ \ \mbox{for}\ \ \frac{1}{2}\leq t\leq T+\frac{1}{2},
$$
$$
D_t=\Omega(t-\frac{1}{2}-T)\ \ \mbox{for}\ \ T+\frac{1}{2}\leq t\leq T+1.
$$
Here $D_{t_0}=D\cap\{t=t_0\}$.

Therefore $\Delta(\tau)$  and  $\Delta(-\tau)$  increase  in size  from
$\tau=\frac{1}{2}$  to $\tau=0$  when  $0\leq t\leq \frac{1}{2}$.  Then they do not move  for  $\frac{1}{2}\leq t\leq T+\frac{1}{2}$.  Note  that $D_t$  consists of
the components $D_t^+$  and $D_t^-$  for 
$\frac{1}{2}\leq t\leq \frac{1}{2}+T$,  where $x_2>0$  in
$D_t^+,\ x_2<0$  in $D_t^-$.  When 
$T+\frac{1}{2}\leq t\leq T+1$  the parts $\Delta(\tau)$  and $\Delta(-\tau)$  
return  back to
the initial position $\tau=\frac{1}{2}$. 

Such moving domain is easy to realize.  We can arrange  that
$V(x,t)=0$  in $D$  for  $0\leq t\leq \frac{1}{2}+\e$  and for
$\frac{1}{2}+T-\e<t\leq T+1, \ V(x,t)=V_1(t)$  in $D_t^+, \
V(x,t)=V_2(t)$ in $D_t^-$  for $\frac{1}{2}\leq t\leq T+\frac{1}{2}$.
Then $E=\frac{\partial V(x,t)}{\partial x}=0$  in $D$.

We consider the Schr\"odinger  equation
(\ref{eq:4.1})  in $D$
with nonzero initial condition
(\ref{eq:4.3})                             
and the zero boundary  condition (\ref{eq:4.2}).

   The gauge  group  $G(\overline D)$  consists  of all  $g(x,t)$  in $\overline D$   
such that $|g(x,t)|=1$.
It follows from the topology  of $D$  that any $g(x,t)\in G(\overline D)$  has the form:
$$
g(x,t)=e^{i n\theta+\frac{i}{h}\varphi(x,t)},
$$
where $\varphi(x,t)$  is 
real-valued and
differentiable in $\overline D$  and $\theta$  is the polar angle  in
$(x_2,t)$-plane.  If $V_1(x,t)$ and $V_2(x,t)$  are gauge equivalent  and if $\gamma$  is
a closed contour in $D$ not homotopic to a point,   then
$$
e\int_\gamma V_1(x,t)dt -e\int_\gamma V_2(x,t)dt
=ih\int_\gamma g^{-1}(x,t)\frac{\partial g}{\partial t}dt
=2\pi h n,\ \ n\in \Z.
$$

Let $\alpha_j=\frac{e}{h} \int_{\frac{1}{2}}^{T+\frac{1}{2}} V_j(t)dt,\ j=1,2,$  and 
suppose $\alpha_1-\alpha_2\neq 2\pi  n, \ \forall n\in\Z$.
Since the electric flux $\alpha=\frac{e}{h}\int_\gamma V(x,t)dt=\alpha_1-\alpha_2\neq 2\pi n,
\ \forall n$  the electric potential $V(x,t)$  is not gauge  equivalent to the zero  potential.                              

Now we are ready to prove Theorem \ref{theo:1.2}.

{\bf Proof of Theorem 1.2}.
Since $u(x,t)$  and $v(x,t)$  have the same initial and boundary conditions (\ref{eq:4.2}),
(\ref{eq:4.3})  and $V(x,t)=0$  for $t\in[0,\frac{1}{2}]$  we have that 
$u(x,t)=v(x,t)$  for $t\in [0,\frac{1}{2}]$.   Denote  by $\Pi_1$  and $\Pi_2$
two connected components of $D\cap(\frac{1}{2},T+\frac{1}{2})$,
$  
\Pi_1=\cup_{\frac{1}{2}\leq t\leq T+\frac{1}{2}}D_t^+,\ \linebreak
\Pi_2=\cup_{\frac{1}{2}\leq t\leq T+\frac{1}{2}}D_t^-.
$
  Note that
in $\Pi_i,i=1,2,$  we have 
$$
u(x,t)=v(x,t)\exp\left(-i\frac{e}{h}\int_{\frac{1}{2}}^tV_i(t')dt'\right),\ 
\ (x,t)\in\Pi_i.
$$
Let 
 $c(x,t)=1$  for  $t\leq \frac{1}{2},
\ c(x,t)=e^{-i\frac{e}{h}\int_{\frac{1}{2}}^tV_i(t')dt'}$  
in $\Pi_i$ 
for $t\in (\frac{1}{2},T+\frac{1}{2}),i=1,2.$  Then
$u(x,t)=c(x,t)v(x,t)$,  i.e.  $V(x,t)$  is gauge equivalent  to zero in 
$D\cap(0,T+\frac{1}{2})$.  
\qed

Note that $u(x,t)$  and $v(x,t)$  satisfy the same equation (\ref{eq:4.5}) in 
\linebreak
$D\cap\{T+\frac{1}{2}\leq t\leq T+1\}$
and 
$$
u(x,T+\frac{1}{2})=e^{-i\alpha_1}v(x,T+\frac{1}{2})\ \ \mbox{in}\ \ D_{T+\frac{1}{2}}^+,
$$
$$
u(x,T+\frac{1}{2})=e^{-i\alpha_2}v(x,T+\frac{1}{2})\ \ \mbox{in}\ \ D_{T+\frac{1}{2}}^-.
$$
We shall show  that the probability densities $|u(x,t)|^2$  and $|v(x,t)|^2$
are not equal identically for $T+\frac{1}{2}< t<T+\frac{1}{2}+\e$.  

To prove that $|u(x,t)|^2\not\equiv |v(x,t)|^2$    we consider 
$w(x,t)=e^{i\alpha_2}u(x,t)$.   Then $w(x,T+\frac{1}{2})=v(x,T+\frac{1}{2})$  
 in $D_{T+\frac{1}{2}}^-$,  
$w(x,T+\frac{1}{2})=e^{i(\alpha_2-\alpha_1)}v(x,T+\frac{1}{2})$ 
in $D_{T+\frac{1}{2}}^+$.
\begin{pro}                                \label{prop:4.1}
Let $(0,x_2^{(0)},T+\frac{1}{2})$  be a point in $D_{T+\frac{1}{2}}^-$  such that
\linebreak
$v(0,x_2^{(0)},T+\frac{1}{2})\neq 0$.
Let $O$  be a small neighborhood of $(0,x_2^{(0)},T+\frac{1}{2})$ in $D$.
Suppose $v(x,T+\frac{1}{2})=w(x,T+\frac{1}{2})$  in $O\cap\{t=T+\frac{1}{2}\}$.
If $|v(x,t)|^2=|w(x,t)|^2$  in $O\cap\{t>T+\frac{1}{2}\}$  
then $v(x,t)=w(x,t)$  in $D\cap\{T+\frac{1}{2}<t<T+\frac{1}{2}+\e\}$.
\end{pro}

{\bf Proof of Proposition \ref{prop:4.1}}.
Let 
$R(x,t)=|v(x,t)|,\ \Phi(x,t)=\arg v(x,t)$,  i.e.
$v(x,t)=R(x,t)e^{i\Phi(x,t)}$.   Substituting in (\ref{eq:4.5})  and separating
the real and the imaginary parts we get
\begin{equation}                             \label{eq:4.8}
-hR_t=\frac{h^2}{2m}(2\nabla R\cdot\nabla \Phi+R\Delta\Phi),
\end{equation}
\begin{equation}                               \label{eq:4.9}
h\Phi_t R=\frac{h^2}{2m}(\Delta R-R|\nabla\Phi|^2).
\end{equation}
Suppose $R$  is given.  Then  (\ref{eq:4.9})  is a first order 
partial differential
equation in $\Phi$  and 
therefore
the initial data $\Phi(x,T+\frac{1}{2})$ 
in $O\cap\{t=T+\frac{1}{2}\}$   uniquely  determines $\Phi(x,t)$  
in the neighborhood $O$.  
Let $w=R_1(x,t)e^{i\Phi_1(x,t)}$.
Note that $w(x,t)$ also satisfies  (\ref{eq:4.5})  
for  $t>T+\frac{1}{2}$  and $\Phi_1$  satisfies (\ref{eq:4.9})
for $t>T+\frac{1}{2}$.
Since $R=R_1$  in $O$  and, since $\Phi_1(x,T+\frac{1}{2})=
\Phi(x,T+\frac{1}{2})$  in $O\cap\{t=T+\frac{1}{2}\}$, we have that 
$w(x,t)=v(x,t)$  in $O$.
Then $w(x,t)=v(x,t)$  for $D\cap (T+\frac{1}{2},T+\frac{1}{2}+\e)$  
by the unique continuation property (see [I], section 6). 
By the continuity in $t$  we get  $v(x,T+\frac{1}{2})=w(x,T+\frac{1}{2})$  
in $D_{T+\frac{1}{2}}$
and  this is a contradiction with 
$w(x,T+\frac{1}{2})=e^{i(\alpha_2-\alpha_1)}v(x,T+\frac{1}{2})$  
in $D_{T+\frac{1}{2}}^+$,  assuming that $v(x,T+\frac{1}{2})\not\equiv 0$ in 
$D_{T+\frac{1}{2}}^+$.  
\qed

Therefore 
$|u(x,t)|^2\not\equiv |v(x,t)|^2$  for  $T+\frac{1}{2}<t<T+\frac{1}{2}+\e$, 
i.e. the AB effect holds.

{\bf Remark 4.1.}  In the proof of Proposition \ref{prop:4.1} we used that 
$R(x,t)=|v(x,t)|\not \equiv 0$  in $D_{T+\frac{1}{2}}^-$
  and $|v(x,t)|\not\equiv 0$  in $D_{T+\frac{1}{2}}^+$.
We  shall  show that  this can be achieved by the appropriate choice  of
the initial condition $u_0(x)$  in (\ref{eq:4.7}).  Choose any 
$v(x,T+\frac{1}{2})$  such 
that $v(x,T+\frac{1}{2})\not\equiv 0$  in $D_{T+\frac{1}{2}}^-$  and  
$v(x,T+\frac{1}{2})\not\equiv 0$  in $D_{T+\frac{1}{2}}^+$.

Solve  the backward initial value  problem  for (\ref{eq:4.5})
with the boundary condition (\ref{eq:4.6}) and the initial  condition 
$v(x,T+\frac{1}{2})$  in
$D_{T+\frac{1}{2}}$.  Then  we take  $v(x,0)$  as the initial condition $u_0(x)$
in (\ref{eq:4.7})  and 
(\ref{eq:4.3}).

Note that the map $u_0(x)\rightarrow v(x,T+\frac{1}{2})$ is an open map of 
$\stackrel{\circ}{H}_1(D_0)\cap H_2(D_0)$  to 
$\stackrel{\circ}{H}_1(D_{T+\frac{1}{2}})\cap H_2(D_{T+\frac{1}{2}})$ (cf., for example, [E1]).
Therefore, $v(x, T+\frac{1}{2})\not\equiv 0$  on $D_{T+\frac{1}{2}}^-$ and
 $v(x, T+\frac{1}{2})\not\equiv 0$  on $D_{T+\frac{1}{2}}^+$    for open dense set of $u_0(x)$,
i.e. the assumption 
in the proof of Proposition \ref{prop:4.1} are satisfied for generic $u_0(x)$.

{\bf Remark 4.2.}
The electric AB effect was studied in [W2] for the electric potentials 
of the form $vV_0(vt,x)$,  where $v$  is a large parameter.  
Note  that the AB effect is an exact physical statement and not an 
asymptotic one.  
The introduction of a large parameter is not justified in this case.

{\bf Acknowledgment.}

Author is grateful to Lev Vaidman of the Physics Department, Tel-Aviv University,
and Eric Hudson of the Physics Department, UCLA,
 for the stimulating discussions. 
I am thankful to Ulf Leonhardt,  St.Andrews University,   for insightful remarks.

Author also grateful to the referees for valuable suggestions that significantly improved the paper.

\end{document}